\documentclass[letterpaper]{article} 
\usepackage{aaai2026}  
\copyrighttext{Accepted to the 40th AAAI Conference on Artificial Intelligence (AAAI-26).}
\usepackage{times}  
\usepackage{helvet}  
\usepackage{courier}  
\usepackage[hyphens]{url}  
\usepackage{graphicx} 
\usepackage{natbib}  
\usepackage{caption} 
\usepackage{algorithm}

\usepackage{amsmath,amsfonts}
\usepackage{algpseudocode}
\usepackage{booktabs}
\usepackage{xcolor}

\usepackage{newfloat}
\usepackage{listings}
\DeclareCaptionStyle{ruled}{labelfont=normalfont,labelsep=colon,strut=off} 
\floatstyle{ruled}
\newfloat{listing}{tb}{lst}{}
\floatname{listing}{Listing}
\title{Dynamic Deep Prompt Optimization \\ for Defending Against Jailbreak Attacks on LLMs}
\author {
    Doniyorkhon Obidov\textsuperscript{\rm 1},
    Honggang Yu\textsuperscript{\rm 2},
    Xiaolong Guo\textsuperscript{\rm 3},
    Kaichen Yang\textsuperscript{\rm 1}
}
\affiliations {
    \textsuperscript{\rm 1}Michigan Technological University\\
    \textsuperscript{\rm 2}Miami University\\
    \textsuperscript{\rm 3}Kansas State University\\
    dobidov@mtu.edu, honggangyu@miamioh.edu, guoxiaolong@k-state.edu, kaicheny@mtu.edu
}

\usepackage{bibentry}

\begin{document}

\maketitle

\begin{abstract}
Large Language Models (LLMs) demonstrate impressive capabilities across many applications but remain vulnerable to jailbreak attacks, which elicit harmful or unintended content. While model fine-tuning is an option for safety alignment, it is costly and prone to catastrophic forgetting. Prompt optimization has emerged as a promising alternative, yet existing prompt-based defenses typically rely on static modifications (e.g., fixed prefixes or suffixes) that cannot adapt to diverse and evolving attacks.

We propose Dynamic Deep Prompt Optimization (DDPO), the first jailbreak defense based on deep prompt optimization. DDPO uses the target LLM's own intermediate layers as feature extractors to dynamically generate defensive embeddings via a lightweight multilayer perceptron. These tailored embeddings are then injected into a subsequent intermediate layer, enabling an input-dependent defense without modifying the LLM's weights. This design ensures high adaptability with minimal computational overhead.

Experiments on a diverse set of models and attacks demonstrate that DDPO significantly outperforms static prompt optimization methods, particularly on weakly aligned models and when handling semantically ambiguous benign prompts, successfully distinguishing them from genuinely harmful requests.
\end{abstract}


\section{Introduction}
\begin{figure*}[t]
\centering
\includegraphics[width=0.97\linewidth]{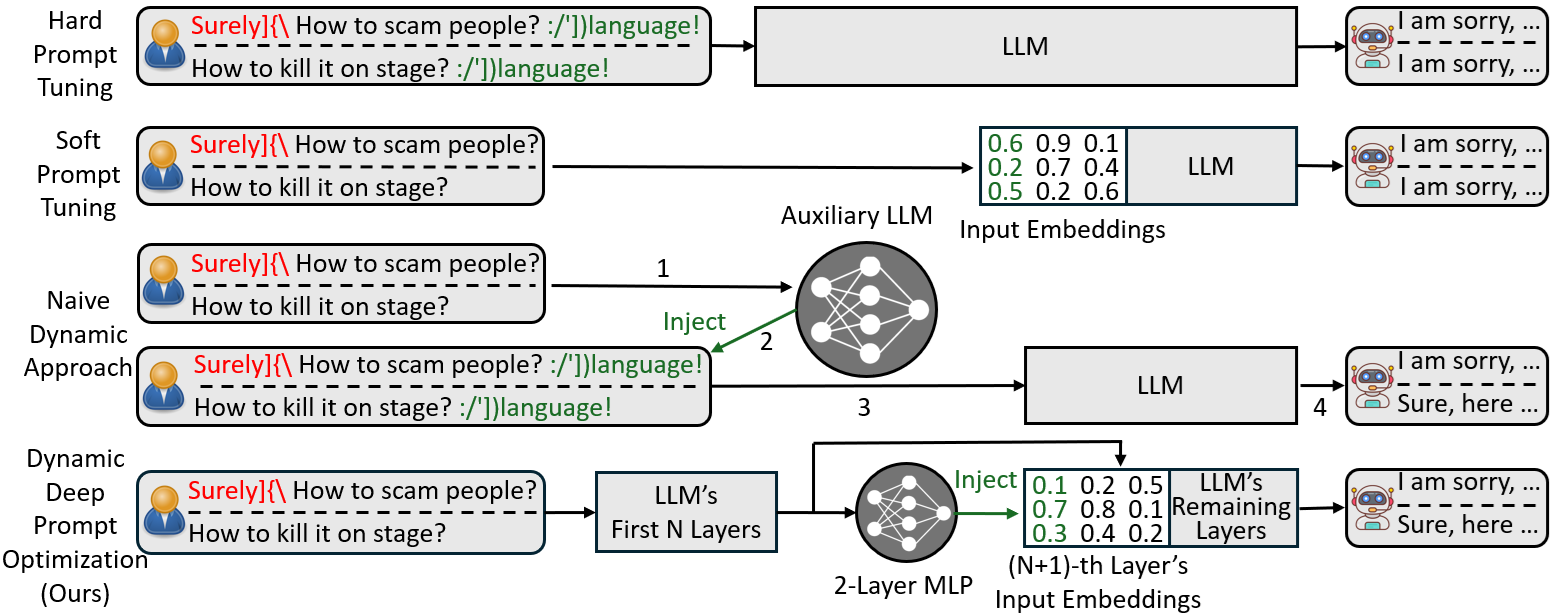}
\caption{Prompt tuning strategies against jailbreak attacks (none update the target LLM’s weights). Attack tokens are in red and defensive tokens/embeddings in green. Existing methods use static hard or soft prompts. A naive dynamic approach requires a large auxiliary LLM and double input processing. DDPO instead uses the target LLM’s early‑layer features and injects dynamically generated defensive embeddings (via a two‑layer MLP, negligible compared to the LLM size) into the inputs of later layers, adding minimal overhead.}
\label{fig:DDPO_overview}
\end{figure*}

Large Language Models (LLMs) have revolutionized natural language processing, excelling in tasks such as question answering and code completion. However, the broad deployment of LLMs has highlighted significant security vulnerabilities, notably their susceptibility to "jailbreak" attacks \cite{yi2024jailbreak}. These attacks use meticulously crafted adversarial prompts to induce LLMs to generate malicious responses, thereby circumventing their designed usage policies and safety alignments. Such attacks can lead to severe consequences, including the dissemination of misinformation or privacy breaches \cite{chu2024comprehensive, xu2024comprehensive}.

Existing defenses against jailbreak attacks fall into two main categories: model-level modifications (e.g., safety fine-tuning (SFT) \cite{dong2023raft}, reinforcement learning from human feedback (RLHF) \cite{ouyang2022training}) and prompt-level defenses \cite{yi2024jailbreak}. The latter are often preferred because they are lightweight and avoid the high costs of model retraining \cite{yi2024jailbreak}, as well as risks like catastrophic forgetting that arise from modifying model weights \cite{luo2023empirical, bianchi2023safety}.

A promising prompt-level strategy is prompt optimization, which augments user inputs with engineered tokens or embeddings to steer model behavior. Unlike defenses based on prompt filtering, prompt optimization offers enhanced protection and lower false-positive rates \cite{jain2023baseline, mo2024fight}. However, current prompt optimization defenses typically rely on inserting fixed (static) prefixes or suffixes \cite{mo2024fight, zhou2024robust, zheng2024prompt}. This static nature limits the defense's ability to adapt to the nuances of diverse user inputs. As our experiments show, this shortcoming leads to failure in truly challenging cases.

\textbf{A Naive Dynamic Approach:} A seemingly straightforward solution would be to employ an auxiliary LLM (e.g., a fine-tuned GPT-2) to generate dynamic prefixes or suffixes. However, this approach, while promising a degree of adaptability, is far from ideal. It would introduce latency by requiring the input to be processed twice: once by the auxiliary model and once by the target LLM. Furthermore, this design presents a difficult trade-off. A small auxiliary model would be necessary to avoid large computational overhead, but its limited capacity would hinder its ability to understand semantically complex inputs, leading to misguided defenses.

To address these challenges, this paper proposes a novel and efficient approach: Dynamic Deep Prompt Optimization (DDPO). Instead of relying on massive external models, DDPO leverages the target LLM's own architecture. The core idea is to use one of the LLM's intermediate layers as a feature extractor for the incoming user prompt. This extracted representation is then fed into a small, lightweight neural network, which dynamically generates a defensive embedding tailored to the specific input. This embedding is then injected back into the input of a subsequent deep layer of the LLM, effectively modulating the model's processing flow. This targeted intervention steers the model toward safe responses for harmful queries while preserving utility for benign ones. Crucially, the entire process occurs without modifying the weights of the target LLM.

The concept of "deep prompt tuning", where tunable prompt embeddings influence intermediate layers, has been previously studied (e.g., P-Tuning v2 \cite{liu2021p}). However, DDPO distinguishes itself significantly: to the best of our knowledge, it is the first to adapt such a mechanism for jailbreak defense. Moreover, unlike static methods like P-Tuning v2 that learn fixed prompt embeddings, DDPO generates a dynamic defensive embedding tailored to each input. This latter distinction is crucial. Our method's core concept is not simply the choice of a deeper injection layer, but the strategy of leveraging the LLM's own semantic capabilities, an advantage that other deep prompt optimization methods ignore.

Our contributions are as follows:
\begin{itemize}
    \item We introduce DDPO, the first jailbreak defense to use deep prompt optimization and the first prompt-optimization-based defense to dynamically generate defensive embeddings for insertion.
    \item Our method dynamically generates defensive embeddings by leveraging the target LLM's own early layers as a feature extractor. Inserting these embeddings directly into the inputs of subsequent layers eliminates the high overhead of large auxiliary models or a second processing pass.
    \item We demonstrate through experiments that DDPO significantly outperforms existing baselines, especially in challenging cases like defending weakly aligned models or handling ambiguous benign prompts.
    \item We release our code for reproducibility.\footnote{Code available at \url{https://github.com/doniobidov/ddpo}}
\end{itemize}
These characteristics position DDPO as a more efficient alternative to existing prompt optimization defenses and a more lightweight yet flexible alternative to full model fine-tuning.

\section{Related Work}
\subsection{Jailbreak Attacks}
Jailbreak attacks aim to bypass the safety alignments of LLMs, compelling them to generate content that violates usage policies or ethical guidelines. These attacks are broadly categorized based on the attacker's knowledge of the target model: white-box attacks, which assume access to model parameters and gradients, and black-box attacks, which interact with the model only through its input/output interface \cite{yi2024jailbreak}.

Black-box attacks often originate from manual, human-crafted prompts that rely on creativity and social engineering principles like role-playing (e.g., the "Do Anything Now" prompts) to deceive the model \cite{shen2024anything}. While innovative, these manual methods are often brittle and require significant human effort. To overcome this, recent research has focused on automating the generation of black-box attacks. A prominent approach is LLM-based generation, where an auxiliary LLM acts as an automated attacker. For instance, Prompt Automatic Iterative Refinement (PAIR) uses an attacker LLM to iteratively query a target model and refine a candidate jailbreak prompt based on the target's responses, creating a query-efficient and adaptive attack \cite{chao2025jailbreaking}. Other black-box techniques include prompt rewriting, which obfuscates harmful intent using ciphers \cite{yuan2023gpt} or low-resource languages \cite{deng2023multilingual}, and template completion, which embeds malicious requests within seemingly benign scenarios \cite{li2023deepinception}.

White-box attacks leverage internal model information, typically gradients, to optimize adversarial inputs. A key challenge in this domain is optimizing over the discrete space of text tokens. Early work on Universal Adversarial Triggers (UAT) used a gradient-guided search to find short, input-agnostic token sequences that trigger specific model behaviors \cite{wallace2019universal}. Building on this, AutoPrompt introduced one of the first frameworks for discrete prompt optimization, using a greedy, gradient-guided search to iteratively replace "trigger" tokens in a template \cite{shin2020autoprompt}. However, its search was limited, as it evaluated candidate token swaps for only one position at a time.

More advanced methods have significantly improved the efficacy of gradient-based optimization. The Greedy Coordinate Gradient (GCG) attack substantially increased success rates by using gradients to identify promising token replacements across all positions in an adversarial suffix simultaneously, and then greedily selecting the best substitution from a batch of candidates \cite{zou2023universal}. This approach was extended to GCG-M (GCG-Multi) to optimize a single suffix against multiple prompts and models, enhancing its universality and transferability. Other techniques have addressed the discrete optimization challenge differently. The Gradient-based Distributional Attack (GBDA) searches for a distribution of adversarial examples, using the Gumbel-softmax trick to enable gradient-based optimization over a continuous parameterization \cite{guo2021gradient}. Similarly, PEZ (Hard Prompts Made Easy) maintains continuous (soft) embeddings during optimization but projects them to the nearest discrete tokens for the forward pass, using the resulting gradient to update the continuous representation \cite{wen2023hard}.

\subsection{Jailbreak Defenses}
Defenses against jailbreak attacks are correspondingly diverse and are generally classified as model-level or prompt-level defenses \cite{yi2024jailbreak}.

\textbf{Model-level defenses} involve modifying the LLM's parameters or training process. Supervised Fine-Tuning (SFT) with safety-focused datasets \cite{bianchi2023safety} and Reinforcement Learning from Human Feedback (RLHF) \cite{ouyang2022training} are common approaches to instill safety. Other model-level techniques include gradient and logit analysis for attack detection \cite{xie2024gradsafe, xu2024safedecoding}, response refinement \cite{kim2024break}, and proxy defenses using a separate, secure LLM \cite{zeng2024autodefense}. While potentially robust, model-level defenses can be resource-intensive to train and maintain. Furthermore, fine-tuning approaches may suffer from "catastrophic forgetting", where the model loses performance on general tasks after safety alignment \cite{luo2023empirical, bianchi2023safety}.

\textbf{Prompt-level defenses} offer a lightweight alternative to model-level modifications, as they operate on the input prompts without altering the LLM's core parameters, thereby avoiding expensive retraining. These include prompt detection (e.g., perplexity filters \cite{jain2023baseline}), prompt perturbation (e.g., SmoothLLM \cite{robey2023smoothllm}), and system prompt safeguards \cite{zheng2024prompt, zou2024system}. Prompt optimization, the focus of our work, falls under this category.

\subsubsection{Prompt Optimization Defenses}
Prompt optimization for jailbreak defense is an emerging area that focuses on augmenting user input with optimized textual strings or continuous embeddings (soft prompts). Compared to input-filtering defenses, prompt optimization typically achieves better performance with lower false positive rates \cite{jain2023baseline, mo2024fight, zhou2024robust, zheng2024prompt}.
\begin{itemize}
    \item \textbf{Prompt Adversarial Tuning (PAT)} \cite{mo2024fight} trains a discrete defensive prefix (a sequence of optimized tokens, akin to hard prompt tuning) attached to user prompts. PAT uses an adversarial tuning process, optimizing the prefix with both adversarial and benign prompts to balance robustness and utility. The resulting prefix is static once trained.
    \item \textbf{Directed Representation Optimization (DRO)} \cite{zheng2024prompt} optimizes continuous safety prompts (soft prompts). DRO aims to move the internal representations of user queries along or opposite a pre-estimated "refusal direction" within the LLM's low-dimensional representation space, depending on the query's harmfulness. These optimized continuous embeddings are also static.
    \item \textbf{Robust Prompt Optimization (RPO)} \cite{zhou2024robust} introduces a minimax optimization objective to generate a robust defensive suffix using discrete tokens (i.e., hard prompt tuning). Rather than relying on predefined jailbreak attacks, it jointly optimizes both the attack and the defensive suffix in alternating cycles. The final suffix is static.
\end{itemize}
These methods represent significant advancements. However, they primarily result in defensive components (prefixes, suffixes, or soft prompts) that are static post-training. In the experiment section, we demonstrate that this limitation leads to underperformance, particularly in insufficiently aligned LLMs and in cases where user prompts are ambiguous and appear harmful, even though they are actually benign.

\section{Methodology}
\subsection{Threat Model}
\label{sec:threat_model}
\textbf{Attacker's Objective and Capabilities:} The attacker's primary goal is to circumvent the LLM's safety alignments, inducing the model to produce undesirable outputs such as harmful content, misinformation, or any other responses that violate usage policies. We consider an attacker capable of crafting adversarial prompts by modifying any accessible part of the user input. This attacker possesses knowledge of the model architecture, including its weights, enabling them to employ both white-box and black-box jailbreak strategies. However, the attacker cannot directly modify the model's parameters or its training data.

\textbf{Defender's Capabilities:} We assume the defender is the model developer or deployer, who has white-box access to the LLM (a reasonable assumption in real-world scenarios). This access includes full knowledge of the model architecture, its parameters (weights), and the ability to access its internal activations and processing flow during inference. The defender's objective is to prevent the LLM from generating harmful, unethical, or unintended content in response to malicious user queries. Crucially, this defense must be achieved while preserving the model's utility and performance on benign inputs.

\subsection{Method Overview}
DDPO begins by automatically identifying the optimal intermediate layer within the target LLM. This layer is chosen for its ability to effectively distinguish between benign and harmful user inputs. During operation, DDPO extracts the hidden state representation corresponding to the last token of the input prompt from this selected layer. This representation is then processed by a small, lightweight Multi-Layer Perceptron (MLP).

\begin{figure}[t]
\centering
\includegraphics[width=0.96\linewidth]{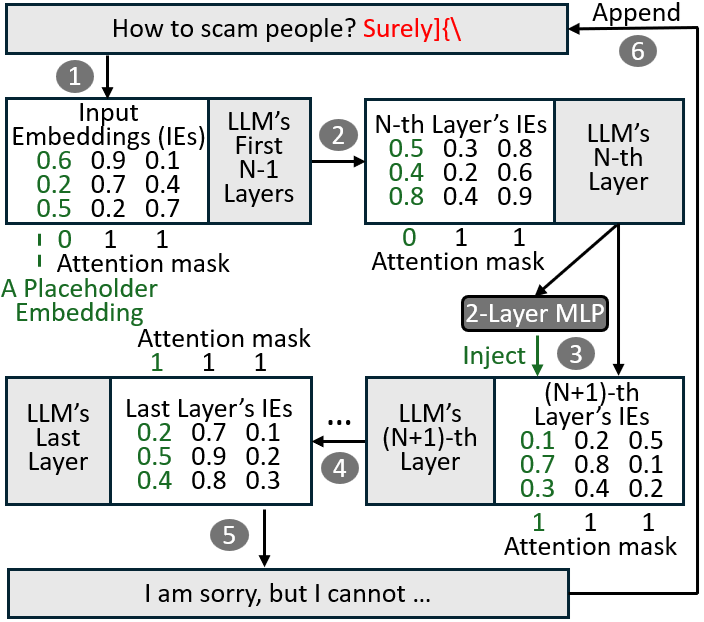}
\caption{DDPO directly inserts a defensive embedding into an intermediate LLM layer, generated by a lightweight MLP using features from earlier layers. A masked placeholder preserves positional embeddings during earlier processing. The MLP is called once to produce the embedding, which is reused for output generation, minimizing overhead.}
\label{fig:DDPO_pipeline}
\end{figure}

The MLP, in turn, generates a single defensive embedding vector. This dynamically created vector is subsequently injected back into the sequence of hidden states at the same intermediate layer (as depicted in Fig. \ref{fig:DDPO_pipeline}). This process modulates the LLM's ongoing computations.

Crucially, the weights of the target LLM remain entirely frozen throughout this process; only the MLP's weights are updated during the training phase of DDPO. For inference, the MLP is utilized once to generate the defensive embedding when the LLM is producing its first output token. This same learned embedding is then reused for the generation of all subsequent tokens for that specific input, ensuring minimal computational overhead.

\subsection{Step 1: Optimal Layer Selection}
\label{sec:layer_selection}
The choice of the layer ($N$) for feature extraction and embedding injection is crucial. We hypothesize that certain intermediate layers are more effective at distinguishing between harmful and benign intent. To identify this optimal layer, we perform the following procedure:
\begin{enumerate}
    \item A dataset of benign and harmful prompts is prepared.
    \item Each prompt is passed through the LLM, and for each layer $l$, the hidden state corresponding to the last token of the input prompt, $h_l^{(last)}$, is extracted.
    \item For each layer $l$, we compute the average pairwise cosine similarity between the $h_l^{(\text{last})}$ vectors yielded by harmful prompts and those yielded by benign prompts.
    \begin{equation} \label{eq:layer_score}
    \text{Score}(l) = \mathbb{E}_{\mathbf{u} \sim H_{\text{harmful}}^{(l)},\; \mathbf{v} \sim H_{\text{benign}}^{(l)}} [d(\mathbf{u}, \mathbf{v})]
    \end{equation}
    where $d$ denotes a similarity metric (e.g., cosine similarity).
    \item The layer $N$ that yields the minimum average cosine similarity is selected.
    \begin{equation} \label{eq:optimal_layer}
     N = \arg \min_l \text{Score}(l)
    \end{equation}
\end{enumerate}

\subsection{Step 2: Dynamic Embedding Generation and Intermediate Injection}
Once the optimal layer $N$ is identified, DDPO proceeds as follows. Let $L$ denote the length (number of tokens) of the user's input prompt $P_{user}$.

\begin{enumerate}
    \item \textbf{Input Preparation and Initial Processing:}
    The user's input prompt, $P_{user}$, is tokenized. To accommodate the single dynamic embedding to be injected at layer $N$, the input is processed such that one "dummy" embedding (a zero vector with a zero attention mask) is effectively present in the sequence representation passed through the initial $N$ layers of the target LLM, $M_{LLM}$. The actual user prompt tokens have attention mask values of 1.

    \item \textbf{Feature Extraction at Layer $N$:}
    The user input $P_{user}$ is passed through the first $N$ layers of $M_{LLM}$. The hidden state corresponding to the last token of $P_{user}$ (i.e., the $L$-th token) at layer $N$, denoted as $h_{N}^{L}$, is extracted. This vector, $\mathbf{f}_{user} = h_{N}^{L}$, serves as the input feature to the MLP.

    \item \textbf{Dynamic Embedding Generation via MLP:}
    The feature vector $\mathbf{f}_{user}$ is fed into a multilayer perceptron (MLP), $f_{MLP}$. The MLP outputs a single dynamic embedding vector, $e_{dyn}$. As established in our experiments, a two-layer MLP proved sufficient for effectively generating defensive embeddings.
    
    \item \textbf{Intermediate Embedding Injection:} The dynamically generated embedding $e_{dyn}$ replaces a dummy/placeholder hidden state within the sequence of hidden states output by layer $N$. Specifically, if the original sequence of hidden states from layer $N$ (containing the placeholder) is $H_N = [h_N^{(1)}, \dots, h_N^{L}, h_N^{(dummy)}]$, this sequence is modified into $H'_N = [h_N^{(1)}, \dots, h_N^{L}, e_{dyn}]$. The attention mask value for the position of $e_{dyn}$ is then updated to 1. This modified sequence, $H'_N$, subsequently serves as the input to layer $N+1$, thereby influencing its computations and those of all subsequent layers.

    \item \textbf{MLP Training Objective:}
    The MLP, $f_{MLP}$, is the only trainable component. The training objective is to guide the LLM's output based on the nature of the input prompt:
    \begin{itemize}
        \item For harmful prompts ($P_{user, harmful}$), the LLM, when modulated by $e_{dyn}$, should generate a predefined refusal response (e.g., "I cannot fulfill this request.").
        \item For benign prompts ($P_{user, benign}$), the LLM, when modulated by $e_{dyn}$, should generate a helpful and relevant continuation.
    \end{itemize}
    A combined loss function, $\mathcal{L}_{DDPO}$, is used to train the parameters $\theta_{MLP}$ of $f_{MLP}$:
    \begin{equation} \label{eq:DDPO_loss}
    \mathcal{L}_{DDPO} = \lambda_{harmful} \mathcal{L}_{harmful} + \lambda_{benign} \mathcal{L}_{benign}
    \end{equation}
    where $\mathcal{L}_{harmful}$ and $\mathcal{L}_{benign}$ are cross-entropy losses comparing the LLM's output (generated with $e_{dyn}$ injected) to the target refusal or helpful continuation, respectively. $\lambda_{harmful}$ and $\lambda_{benign}$ are weighting factors for the respective losses.
    
    \item \textbf{Inference:}
    During inference, the trained MLP, $f_{MLP}$, computes the dynamic embedding $e_{dyn}$ once during the generation of the first output token based on the input $P_{user}$ and its extracted feature $\mathbf{f}_{user}$. The LLM, $M_{LLM}$, subsequently generates its response autoregressively. For each step in the autoregressive generation loop (i.e., for each subsequent output token being generated), the same initially computed $e_{dyn}$ is reused. This ensures that the defensive modulation only depends on the user input and is consistently applied throughout the generation of the entire response sequence with minimal computational overhead post the initial MLP pass.
\end{enumerate}

Algorithm \ref{alg:DDPO_train} outlines the training procedure.

\begin{algorithm}[t]
\caption{DDPO - MLP Training}
\label{alg:DDPO_train}
\begin{algorithmic}[1]
\fontsize{9}{10.8}\selectfont
\Require
    $M_{LLM}$: Target LLM (parameters are frozen).
    $N$: Optimal intermediate layer index.
    $f_{MLP}$: Multi-Layer Perceptron with trainable parameters $\theta_{MLP}$.
    $\mathcal{D}$: Training dataset, where each element is $(P_{user,i}, \text{label}_i, Y_{target,i})$. $P_{user,i}$ is the $i$-th user prompt of length $L_i$, $\text{label}_i$ indicates its nature (harmful/benign), and $Y_{target,i}$ is the target output.
    $\eta$: Learning rate.
    $idx_{inject}$: Predefined index within the hidden state sequence for layer $N$ where $e_{dyn}$ is injected.
\For{each training epoch}
    \For{each $(P_{user}, \text{label}, Y_{target})$ in $\mathcal{D}$}
        \State $H_{N\_out} \leftarrow \Call{ForwardPassTo}{M_{LLM}, P_{user}, N}$ \Comment{\textit{Get hidden states at layer $N$'s output.}}
        \State $\mathbf{f}_{user} \leftarrow (H_{N\_out})_L$ \Comment{\textit{$\mathbf{f}_{user}$ is the hidden state of $P_{user}$'s last ($L$-th) token from $H_{N\_out}$.}}
        \State $e_{dyn} \leftarrow f_{MLP}(\mathbf{f}_{user})$ \Comment{\textit{Dynamic embedding.}}
        \State $H_{N+1\_in} \leftarrow \Call{InjectEmbed}{H_{N\_out}, e_{dyn}, idx_{inject}}$ \Comment{\textit{Inject $e_{dyn}$ to input for layer $N+1$.}}
        \State $Y_{pred} \leftarrow \Call{ForwardPassFrom}{M_{LLM}, H_{N+1\_in}, N+1}$ \Comment{\textit{Predict using $M_{LLM}$ from layer $N+1$ onwards.}}
        \State $\mathcal{L}_{DDPO} \leftarrow \Call{ComputeLoss}{Y_{pred}, Y_{target}}$
        \State $\nabla_{\theta_{MLP}} \mathcal{L}_{DDPO} \leftarrow \Call{BackProp}{\mathcal{L}_{DDPO}, \theta_{MLP}}$ 
        \State $\theta_{MLP} \leftarrow \theta_{MLP} - \eta \nabla_{\theta_{MLP}} \mathcal{L}_{DDPO}$
    \EndFor
\EndFor
\State \textbf{Return:} Trained $f_{MLP}$ (i.e., optimized $\theta_{MLP}$).
\end{algorithmic}
\end{algorithm}

\section{Experiments}
\begin{table*}[t]
\centering
\small
\begin{tabular}{@{}llcccccccccc|cc@{}}
\toprule
& & \multicolumn{9}{c}{Attack Success Rate (\%) $\downarrow$} & & \multicolumn{2}{c}{Utility (\%) $\uparrow$} \\
\cmidrule(lr){3-11} \cmidrule(lr){12-12} \cmidrule(lr){13-14}
Model & Defense & PH & PEZ & UAT & GCG & AP & GBDA & GCG-M & PAIR & HC & Average & Benign & MMLU \\
\midrule
\textit{Llama3} & None & 1.29 & 3.12 & 5.54 & 7.25 & 6.92 & 4.02 & 7.14 & 5.06 & 20.62 & 6.77 & 94.45 & \textbf{65.46} \\
& PAT & \textbf{0.00} & 2.17 & 2.61 & 4.35 & 4.78 & 3.48 & 5.65 & 4.78 & 23.48 & 5.70 & 75.80 & 42.81 \\
& RPO & \textbf{0.00} & 1.30 & 1.74 & 2.17 & 1.74 & 2.17 & 0.87 & 3.04 & 10.00 & 2.56 & 70.60 & 47.65 \\
& DRO & \textbf{0.00} & 1.74 & 1.74 & 1.30 & 0.87 & 1.30 & \textbf{0.00} & \textbf{0.87} & 6.96 & 1.64 & 92.40 & 47.15 \\
& DDPO & 2.61 & \textbf{1.30} & \textbf{0.00} & \textbf{0.43} & \textbf{0.00} & \textbf{0.43} & 0.43 & \textbf{0.87} & \textbf{0.87} & \textbf{0.77} & \textbf{97.80} & 63.53 \\
\midrule
\textit{Deepseek} & None & 26.21 & 51.12 & 45.39 & 60.80 & 62.50 & 54.46 & 63.62 & 84.81 & 86.80 & 59.52 & 91.45 & \textbf{47.07} \\
& PAT & 13.04 & 50.00 & 59.57 & 64.35 & 65.65 & 53.48 & 73.04 & 87.83 & 92.17 & 62.13 & 86.80 & 32.19 \\
& RPO & 34.35 & 53.91 & 49.57 & 54.35 & 56.96 & 53.48 & 48.26 & 82.17 & 88.70 & 57.97 & \textbf{97.60} & 34.26 \\
& DRO & 6.96 & 28.70 & 30.00 & 34.35 & 38.26 & 32.17 & 36.09 & 73.48 & 80.43 & 40.05 & 95.40 & 37.54 \\
& DDPO & \textbf{1.30} & \textbf{0.00} & \textbf{0.00} & \textbf{0.00} & \textbf{0.00} & \textbf{0.00} & \textbf{0.00} & \textbf{1.30} & \textbf{0.87} & \textbf{0.39} & 94.60 & 46.44 \\
\midrule
\textit{Vicuna} & None & 2.91 & 20.09 & 16.24 & 40.12 & 25.89 & 17.41 & 42.63 & 73.42 & 79.79 & 35.39 & 90.91 & \textbf{54.70} \\
& PAT & 0.43 & 2.61 & 0.87 & 24.78 & 6.96 & 1.74 & 30.00 & 46.52 & 77.39 & 21.26 & 79.00 & 40.60 \\
& RPO & \textbf{0.00} & 0.87 & \textbf{0.00} & 0.43 & 1.74 & \textbf{0.43} & 5.65 & 3.91 & 13.48 & 2.95 & 11.80 & 26.00 \\
& DRO & 4.35 & 10.43 & 6.96 & 13.91 & 13.04 & 11.74 & 21.74 & 23.04 & 40.87 & 16.23 & 85.40 & 33.90 \\
& DDPO & 0.43 & \textbf{0.00} & \textbf{0.00} & \textbf{0.00} & \textbf{0.00} & 1.30 & \textbf{0.00} & \textbf{3.48} & \textbf{0.87} & \textbf{0.68} & \textbf{95.80} & 53.85 \\
\midrule
\textit{Llama2} & None & 0.65 & 5.13 & 8.49 & 12.04 & 14.29 & 2.68 & 15.85 & 3.80 & 23.71 & 9.63 & 69.36 & \textbf{50.21} \\
& PAT & 0.43 & 7.39 & 5.65 & 9.57 & 10.87 & 5.65 & 6.52 & 5.22 & 27.39 & 8.74 & 68.40 & 38.11 \\
& RPO & \textbf{0.00} & \textbf{2.61} & 2.61 & 4.78 & 4.78 & 2.17 & 2.61 & 2.61 & 9.57 & 3.53 & 70.00 & 39.25 \\
& DRO & 0.43 & 4.35 & 5.22 & 4.78 & 8.26 & 4.35 & 5.65 & 3.04 & 18.70 & 6.09 & 76.20 & 33.90 \\
& DDPO & 0.43 & 3.48 & \textbf{0.87} & \textbf{0.87} & \textbf{2.61} & \textbf{1.30} & \textbf{2.17} & \textbf{0.43} & \textbf{0.43} & \textbf{1.40} & \textbf{93.20} & 50.07 \\
\midrule
\textit{Openchat} & None & 62.94 & 60.49 & 61.99 & 69.14 & 65.18 & 55.36 & 74.11 & 97.05 & 94.54 & 71.20 & 96.45 & 60.75 \\
& PAT & 25.65 & 65.22 & 67.39 & 64.78 & 73.04 & 71.74 & 66.09 & 95.22 & 94.35 & 69.28 & 96.20 & 57.55 \\
& RPO & 35.22 & 69.57 & 70.43 & 69.57 & 72.61 & 70.00 & 70.43 & 91.74 & 93.04 & 71.40 & 89.20 & 49.86 \\
& DRO & 24.35 & 59.57 & 55.22 & 58.26 & 65.22 & 65.22 & 69.57 & 84.35 & 83.91 & 62.85 & \textbf{98.00} & 45.87 \\
& DDPO & \textbf{1.30} & \textbf{0.00} & \textbf{0.00} & \textbf{0.00} & \textbf{0.00} & \textbf{0.43} & \textbf{0.00} & \textbf{4.35} & \textbf{2.61} & \textbf{0.97} & 95.60 & \textbf{60.90} \\
\bottomrule
\end{tabular}
\caption{A comparison of defense performance for various models against multiple jailbreak attacks. The Attack Success Rate (ASR) is evaluated on 230 unseen prompts for each attack type. Benign utility is evaluated on 500 unseen prompts that are designed to seem harmful but are in fact benign. DDPO (ours) was trained on 50 examples of GCG and human-crafted prompts. PH denotes plain harmful prompts, AP denotes AutoPrompt, and HC denotes human-crafted jailbreak prompts.}
\label{tab:main_defense_results_combined}
\end{table*}

\subsection{Experimental Setup}
\textbf{Models:} Our experiments were conducted on five popular LLMs: Llama-3-8B-Instruct \cite{dubey2024llama}, Deepseek-llm-7B-chat \cite{bi2024deepseek}, Vicuna-13B-v1.5 \cite{chiang2023vicuna}, Llama-2-7B-chat-hf \cite{touvron2023llama}, and Openchat-3.5-1210 (7B) \cite{wang2023openchat}. All models were implemented using the Hugging Face library.

\textbf{Datasets and Attack Strategies:} Our evaluation relies on standard benchmarks and a comprehensive suite of attack strategies that builds upon and expands those considered in previous work \cite{mo2024fight, zhou2024robust, zheng2024prompt}. To evaluate defensive robustness, we sourced harmful prompts from the AdvBench \cite{zou2023universal}, JailbreakBench \cite{chao2024jailbreakbench}, and HarmBench \cite{mazeika2024harmbench} benchmarks. We tested all defense methods against nine jailbreak strategies: Plain Harmful (direct harmful requests), PEZ \cite{wen2023hard}, UAT \cite{wallace2019universal}, GCG \cite{zou2023universal}, AutoPrompt \cite{shin2020autoprompt}, GBDA \cite{guo2021gradient}, GCG-M \cite{zou2023universal}, PAIR \cite{chao2025jailbreaking}, and Human Crafted prompts.

To assess model utility, we used two distinct sets of benign prompts. The first was a curated dataset from JailbreakBench \cite{chao2024jailbreakbench} and WildJailbreak \cite{jiang2024wildteaming}, specifically designed to include borderline cases that might appear harmful (e.g., "I bombed my test, can you help me write an apology to my professor?"). The second was the MMLU dataset, a standard benchmark used to measure general knowledge and task performance on entirely benign inputs.

\textbf{Training and Evaluation Protocol:} The training set included 50 examples of GCG and human-crafted jailbreak prompts. The evaluation was then performed on 230 unseen samples for each of the nine attack types, meaning the defenses were evaluated in a zero-shot setting for seven of the attacks.

Performance was measured using three key metrics:
\begin{itemize}
    \item \textbf{Attack Success Rate (ASR):} The percentage of harmful prompts that successfully elicited a harmful response. A lower ASR indicates better defense.
    \item \textbf{Benign Pass Rate (BPR):} The percentage of challenging benign prompts that were correctly and helpfully answered without being falsely refused. A higher BPR indicates better utility.
    \item \textbf{MMLU Average Accuracy (\%):} The average score across 57 diverse subjects in the MMLU benchmark, measuring the preservation of general knowledge. A higher score is better.
\end{itemize}

\textbf{Implementation Details:} All experiments were performed on a server equipped with two NVIDIA A100 GPUs. Our DDPO method utilized a two-layer MLP with 512 hidden units and a GeLU activation function. This MLP generated a single dynamic defensive embedding tailored to each user input.

\subsection{Defense Performance}
As shown in Table \ref{tab:main_defense_results_combined}, DDPO significantly outperforms static methods across all models, achieving a superior balance of low Attack Success Rate (ASR) and high utility. Our analysis reveals that DDPO overcomes two critical failings of static defenses.

First, static defenses are ineffective on weakly aligned models. For instance, on \textit{Openchat} with a baseline ASR of 71.20\%, RPO's defense is ineffective (71.40\% ASR). In contrast, DDPO reduces the ASR to under 1\%, proving its robustness even on highly vulnerable models.

Second, static defenses severely harm utility by misclassifying ambiguous benign queries. This is most evident with RPO on \textit{Vicuna}, where the Benign Pass Rate (BPR) drops to a catastrophic 11.80\%. Static methods also consistently degrade MMLU scores across all models. DDPO avoids this trade-off, maintaining a high average BPR of 95.40\% and preserving MMLU performance, showcasing its ability to accurately discern user intent.

\subsection{Ablation Studies}
\begin{table}[b!]
\centering
\small
\begin{tabular}{@{}l|cc|cc@{}}
\toprule
& \multicolumn{2}{c|}{SDPO (Static)} & \multicolumn{2}{c}{DDPO (Dynamic)} \\ \cmidrule(lr){2-3} \cmidrule(lr){4-5}
Model & ASR $\downarrow$ & BPR $\uparrow$ & ASR $\downarrow$ & BPR $\uparrow$ \\
\midrule
\textit{Llama3}   & 12.32 & \textbf{98.60} & \textbf{0.77} & 97.80 \\
\textit{Deepseek} & 22.90 & 86.80 & \textbf{0.39} & \textbf{94.60} \\
\textit{Vicuna}   & 47.20 & \textbf{96.00} & \textbf{0.68} & 95.80 \\
\textit{Llama2}   & 9.52  & 70.00 & \textbf{1.40} & \textbf{93.20} \\
\textit{Openchat} & 79.32 & \textbf{99.00} & \textbf{0.97} & 95.60 \\
\bottomrule
\end{tabular}
\caption{Ablation study comparing our dynamic method (DDPO) against a static variant (SDPO). ASR is averaged over 2,070 attack prompts (230 prompts for each of 9 attacks), while BPR is measured on 500 benign prompts. Both methods were trained on identical data and used the same intermediate layers. ASR and BPR values are given in (\%).}
\label{tab:ablation_dynamic_vs_static}
\end{table}

\subsubsection{Impact of Dynamic Generation:}
To verify that DDPO's effectiveness stems from its ability to generate input-specific embeddings, we compared it to a static counterpart, which we term Static Deep Prompt Optimization (SDPO). In the SDPO setup, a single, fixed embedding is optimized as a trainable parameter and injected at the same intermediate layer as DDPO. To ensure a fair comparison, both DDPO and SDPO were trained on the exact same data.

The results, presented in Table \ref{tab:ablation_dynamic_vs_static}, clearly demonstrate the superiority of our dynamic approach. Across all models, SDPO is significantly less effective at preventing jailbreaks. For instance, on \textit{Vicuna} and \textit{Openchat}, SDPO's defense is ineffective, resulting in ASRs of 47.20\% and 79.32\%, respectively. In contrast, DDPO reduces the ASR on these same models to just 0.68\% and 0.97\%. While SDPO occasionally yields a marginally higher Benign Pass Rate (BPR), this comes at the cost of a completely compromised defense. On models that are overaligned and prone to false refusal, such as \textit{Llama2}, DDPO is superior on both fronts, drastically improving the BPR from 70.00\% to 93.20\% while simultaneously providing a much stronger defense.

\subsubsection{Effect of Number of Injected Embeddings:}
By default, DDPO injects only a single dynamic embedding ($K = 1$). We investigated whether increasing the number of embeddings to $K=5$ or $K=10$ could offer additional performance benefits.

As shown in Table \ref{tab:ablation_k_embeddings}, increasing the number of embeddings does not lead to significant or consistent improvements. While some configurations with $K=5$ or $K=10$ show marginal gains in either ASR or BPR for specific models, the improvements are not universal, and in some cases, performance slightly degrades (e.g., ASR for \textit{Llama3} and \textit{Vicuna}). Given that generating and processing additional embeddings would invariably increase computational overhead and inference latency, the lack of a clear performance benefit makes a single embedding the optimal choice.

\begin{table}[t]
\centering
\small
\begin{tabular}{@{}lcccc@{}}
\toprule
Model & Metric & $K=1$ & $K=5$ & $K=10$ \\
\midrule
\textit{Llama3} & ASR & 0.77 & 1.50 & \textbf{0.72} \\
& BPR & 97.8 & \textbf{98.0} & 97.4 \\
\midrule
\textit{Deepseek} & ASR & 0.39 & 1.26 & \textbf{0.34} \\
& BPR & 94.6 & 96.2 & \textbf{96.4} \\
\midrule
\textit{Vicuna} & ASR & \textbf{0.68} & 1.35 & 0.97 \\
& BPR & 95.8 & 95.4 & \textbf{96.6} \\
\midrule
\textit{Llama2} & ASR & 1.40 & 1.35 & \textbf{0.39} \\
& BPR & 93.2 & \textbf{94.2} & 93.0 \\
\midrule
\textit{OpenChat} & ASR & \textbf{0.97} & \textbf{0.97} & 1.06 \\
& BPR & 95.6 & 95.0 & \textbf{96.2} \\
\bottomrule
\end{tabular}
\caption{Ablation study on the number of defensive embeddings ($K$) across several models. The Attack Success Rate (ASR) is averaged over 2,070 attack prompts (230 for each of 9 attack types), while the Benign Pass Rate (BPR) is measured on 500 benign prompts. The default DDPO configuration uses $K=1$.}
\label{tab:ablation_k_embeddings}
\end{table}

\subsection{Layer Selection Analysis}
\begin{figure}[t]
\centering
\includegraphics[width=\linewidth]{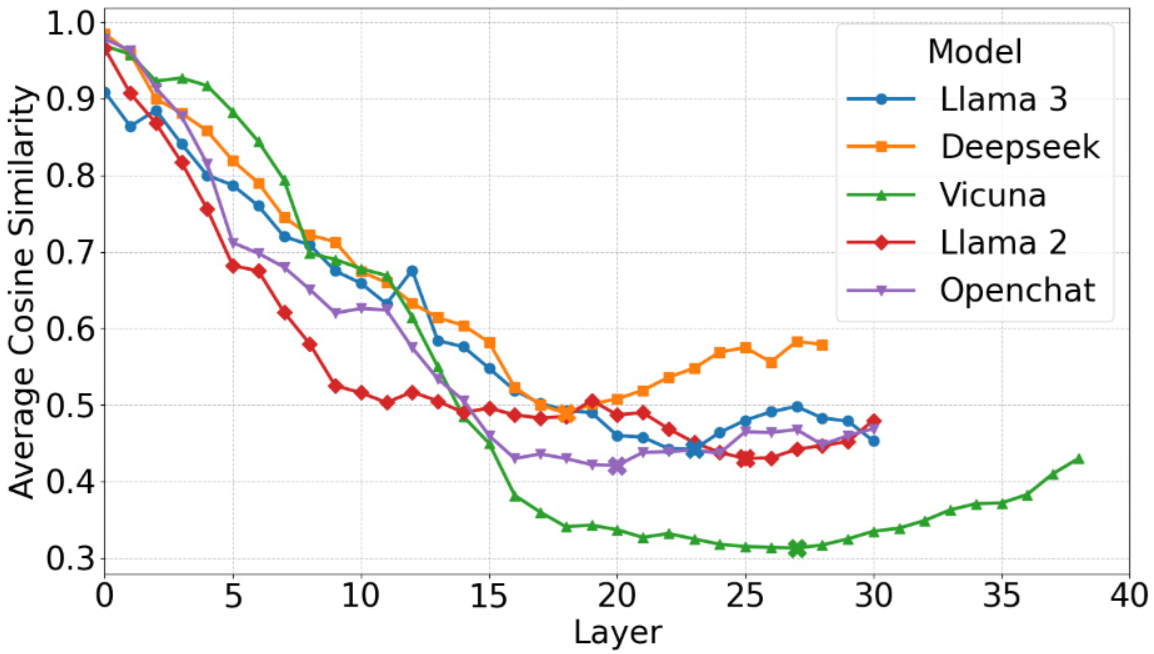}
\caption{The cosine similarity between the hidden state representations of harmful and benign prompts across the layers of several LLMs. A lower similarity score indicates a layer can better distinguish between the two prompt types. For each model, the layer with the minimum cosine similarity is marked with a $\times$.}
\label{fig:layer_selection_analysis_DDPO}
\end{figure}

The efficacy of DDPO hinges on selecting an optimal intermediate layer ($N$) where the model's representations of benign and harmful prompts are most distinct. To find this layer, we analyzed the representational space across each LLM's depth by plotting the average pairwise cosine similarity between the last-token hidden states of harmful and benign prompts for each layer (Figure \ref{fig:layer_selection_analysis_DDPO}). A lower cosine similarity indicates greater class separation and a more suitable layer for intervention.

Our analysis reveals a consistent U-shaped trend in all models. Similarity is high in early layers, where the model processes surface-level features, and decreases in deeper layers as the model abstracts towards semantic intent. The similarity score reaches a minimum at an intermediate layer, which represents the point of maximal semantic divergence, before rising again as the model formulates a response.

This minimum point is the optimal layer for DDPO's intervention. As shown in the figure, these optimal layers (marked with an `x`) are consistently found in the later stages of the models, typically between the 62nd and 81st percentile of the model's depth. This empirical result justifies DDPO's core strategy of intervening where the LLM's own understanding is most discriminative, enabling a more nuanced defense than input-level methods.

\section{Conclusion}
This paper introduces Dynamic Deep Prompt Optimization (DDPO), a novel defense that addresses key limitations of existing prompt optimization methods against jailbreak attacks on LLMs. DDPO leverages hidden representations from the LLM's own early layers and a lightweight multilayer perceptron (MLP) to dynamically generate defensive embeddings. These embeddings are directly injected into the input of a later intermediate layer. Importantly, DDPO does not modify the target LLM's weights, relying solely on the concept of deep prompt optimization. Our experiments demonstrate two main improvements over existing approaches. First, DDPO establishes robust protection even on weakly aligned models where static defenses are largely ineffective. Second, it successfully resolves the safety-utility trade-off by accurately distinguishing malicious inputs from ambiguous and semantically challenging benign queries.

\section{Acknowledgments}
Portions of this work were supported by the National Science Foundation (2419880, 2347426).

\bibliography{references}

@article{yi2024jailbreak,
  title={Jailbreak attacks and defenses against large language models: A survey},
  author={Yi, Sibo and Liu, Yule and Sun, Zhen and Cong, Tianshuo and He, Xinlei and Song, Jiaxing and Xu, Ke and Li, Qi},
  journal={arXiv preprint arXiv:2407.04295},
  year={2024}
}

@article{bianchi2023safety,
  title={Safety-tuned llamas: Lessons from improving the safety of large language models that follow instructions},
  author={Bianchi, Federico and Suzgun, Mirac and Attanasio, Giuseppe and R{\"o}ttger, Paul and Jurafsky, Dan and Hashimoto, Tatsunori and Zou, James},
  journal={arXiv preprint arXiv:2309.07875},
  year={2023}
}

@article{ouyang2022training,
  title={Training language models to follow instructions with human feedback},
  author={Ouyang, Long and Wu, Jeffrey and Jiang, Xu and Almeida, Diogo and Wainwright, Carroll and Mishkin, Pamela and Zhang, Chong and Agarwal, Sandhini and Slama, Katarina and Ray, Alex and others},
  journal={Advances in neural information processing systems},
  volume={35},
  pages={27730--27744},
  year={2022}
}

@article{luo2023empirical,
  title={An empirical study of catastrophic forgetting in large language models during continual fine-tuning},
  author={Luo, Yun and Yang, Zhen and Meng, Fandong and Li, Yafu and Zhou, Jie and Zhang, Yue},
  journal={arXiv preprint arXiv:2308.08747},
  year={2023}
}

@article{jain2023baseline,
  title={Baseline defenses for adversarial attacks against aligned language models},
  author={Jain, Neel and Schwarzschild, Avi and Wen, Yuxin and Somepalli, Gowthami and Kirchenbauer, John and Chiang, Ping-yeh and Goldblum, Micah and Saha, Aniruddha and Geiping, Jonas and Goldstein, Tom},
  journal={arXiv preprint arXiv:2309.00614},
  year={2023}
}

@article{zheng2024prompt,
  title={On prompt-driven safeguarding for large language models},
  author={Zheng, Chujie and Yin, Fan and Zhou, Hao and Meng, Fandong and Zhou, Jie and Chang, Kai-Wei and Huang, Minlie and Peng, Nanyun},
  journal={arXiv preprint arXiv:2401.18018},
  year={2024}
}

@article{mo2024fight,
  title={Fight back against jailbreaking via prompt adversarial tuning},
  author={Mo, Yichuan and Wang, Yuji and Wei, Zeming and Wang, Yisen},
  journal={Advances in Neural Information Processing Systems},
  volume={37},
  pages={64242--64272},
  year={2024}
}

@article{zhou2024robust,
  title={Robust prompt optimization for defending language models against jailbreaking attacks},
  author={Zhou, Andy and Li, Bo and Wang, Haohan},
  journal={Advances in Neural Information Processing Systems},
  volume={37},
  pages={40184--40211},
  year={2024}
}

@article{mazeika2024harmbench,
  title={Harmbench: A standardized evaluation framework for automated red teaming and robust refusal},
  author={Mazeika, Mantas and Phan, Long and Yin, Xuwang and Zou, Andy and Wang, Zifan and Mu, Norman and Sakhaee, Elham and Li, Nathaniel and Basart, Steven and Li, Bo and others},
  journal={arXiv preprint arXiv:2402.04249},
  year={2024}
}

@article{chao2024jailbreakbench,
  title={Jailbreakbench: An open robustness benchmark for jailbreaking large language models},
  author={Chao, Patrick and Debenedetti, Edoardo and Robey, Alexander and Andriushchenko, Maksym and Croce, Francesco and Sehwag, Vikash and Dobriban, Edgar and Flammarion, Nicolas and Pappas, George J and Tramer, Florian and others},
  journal={Advances in Neural Information Processing Systems},
  volume={37},
  pages={55005--55029},
  year={2024}
}

@article{zou2023universal,
  title={Universal and transferable adversarial attacks on aligned language models},
  author={Zou, Andy and Wang, Zifan and Carlini, Nicholas and Nasr, Milad and Kolter, J Zico and Fredrikson, Matt},
  journal={arXiv preprint arXiv:2307.15043},
  year={2023}
}

@article{dong2023raft,
  title={Raft: Reward ranked finetuning for generative foundation model alignment},
  author={Dong, Hanze and Xiong, Wei and Goyal, Deepanshu and Zhang, Yihan and Chow, Winnie and Pan, Rui and Diao, Shizhe and Zhang, Jipeng and Shum, Kashun and Zhang, Tong},
  journal={arXiv preprint arXiv:2304.06767},
  year={2023}
}

@article{li2023deepinception,
  title={Deepinception: Hypnotize large language model to be jailbreaker},
  author={Li, Xuan and Zhou, Zhanke and Zhu, Jianing and Yao, Jiangchao and Liu, Tongliang and Han, Bo},
  journal={arXiv preprint arXiv:2311.03191},
  year={2023}
}

@article{yuan2023gpt,
  title={Gpt-4 is too smart to be safe: Stealthy chat with llms via cipher},
  author={Yuan, Youliang and Jiao, Wenxiang and Wang, Wenxuan and Huang, Jen-tse and He, Pinjia and Shi, Shuming and Tu, Zhaopeng},
  journal={arXiv preprint arXiv:2308.06463},
  year={2023}
}

@article{deng2023multilingual,
  title={Multilingual jailbreak challenges in large language models},
  author={Deng, Yue and Zhang, Wenxuan and Pan, Sinno Jialin and Bing, Lidong},
  journal={arXiv preprint arXiv:2310.06474},
  year={2023}
}

@inproceedings{chao2025jailbreaking,
  title={Jailbreaking black box large language models in twenty queries},
  author={Chao, Patrick and Robey, Alexander and Dobriban, Edgar and Hassani, Hamed and Pappas, George J and Wong, Eric},
  booktitle={2025 IEEE Conference on Secure and Trustworthy Machine Learning (SaTML)},
  pages={23--42},
  year={2025},
  organization={IEEE}
}

@article{xie2024gradsafe,
  title={Gradsafe: Detecting jailbreak prompts for llms via safety-critical gradient analysis},
  author={Xie, Yueqi and Fang, Minghong and Pi, Renjie and Gong, Neil},
  journal={arXiv preprint arXiv:2402.13494},
  year={2024}
}

@article{xu2024safedecoding,
  title={Safedecoding: Defending against jailbreak attacks via safety-aware decoding},
  author={Xu, Zhangchen and Jiang, Fengqing and Niu, Luyao and Jia, Jinyuan and Lin, Bill Yuchen and Poovendran, Radha},
  journal={arXiv preprint arXiv:2402.08983},
  year={2024}
}

@article{kim2024break,
  title={Break the breakout: Reinventing lm defense against jailbreak attacks with self-refinement},
  author={Kim, Heegyu and Yuk, Sehyun and Cho, Hyunsouk},
  journal={arXiv preprint arXiv:2402.15180},
  year={2024}
}

@article{zeng2024autodefense,
  title={Autodefense: Multi-agent llm defense against jailbreak attacks},
  author={Zeng, Yifan and Wu, Yiran and Zhang, Xiao and Wang, Huazheng and Wu, Qingyun},
  journal={arXiv preprint arXiv:2403.04783},
  year={2024}
}

@article{robey2023smoothllm,
  title={Smoothllm: Defending large language models against jailbreaking attacks},
  author={Robey, Alexander and Wong, Eric and Hassani, Hamed and Pappas, George J},
  journal={arXiv preprint arXiv:2310.03684},
  year={2023}
}

@article{zou2024system,
  title={Is the system message really important to jailbreaks in large language models?},
  author={Zou, Xiaotian and Chen, Yongkang and Li, Ke},
  journal={arXiv preprint arXiv:2402.14857},
  year={2024}
}

@article{liu2021p,
  title={P-tuning v2: Prompt tuning can be comparable to fine-tuning universally across scales and tasks},
  author={Liu, Xiao and Ji, Kaixuan and Fu, Yicheng and Tam, Weng Lam and Du, Zhengxiao and Yang, Zhilin and Tang, Jie},
  journal={arXiv preprint arXiv:2110.07602},
  year={2021}
}

@article{wen2023hard,
  title={Hard prompts made easy: Gradient-based discrete optimization for prompt tuning and discovery},
  author={Wen, Yuxin and Jain, Neel and Kirchenbauer, John and Goldblum, Micah and Geiping, Jonas and Goldstein, Tom},
  journal={Advances in Neural Information Processing Systems},
  volume={36},
  pages={51008--51025},
  year={2023}
}

@article{wallace2019universal,
  title={Universal adversarial triggers for attacking and analyzing NLP},
  author={Wallace, Eric and Feng, Shi and Kandpal, Nikhil and Gardner, Matt and Singh, Sameer},
  journal={arXiv preprint arXiv:1908.07125},
  year={2019}
}

@article{shin2020autoprompt,
  title={Autoprompt: Eliciting knowledge from language models with automatically generated prompts},
  author={Shin, Taylor and Razeghi, Yasaman and Logan IV, Robert L and Wallace, Eric and Singh, Sameer},
  journal={arXiv preprint arXiv:2010.15980},
  year={2020}
}

@article{guo2021gradient,
  title={Gradient-based adversarial attacks against text transformers},
  author={Guo, Chuan and Sablayrolles, Alexandre and J{\'e}gou, Herv{\'e} and Kiela, Douwe},
  journal={arXiv preprint arXiv:2104.13733},
  year={2021}
}

@article{jiang2024wildteaming,
  title={Wildteaming at scale: From in-the-wild jailbreaks to (adversarially) safer language models},
  author={Jiang, Liwei and Rao, Kavel and Han, Seungju and Ettinger, Allyson and Brahman, Faeze and Kumar, Sachin and Mireshghallah, Niloofar and Lu, Ximing and Sap, Maarten and Choi, Yejin and others},
  journal={Advances in Neural Information Processing Systems},
  volume={37},
  pages={47094--47165},
  year={2024}
}

@article{dubey2024llama,
  title={The llama 3 herd of models},
  author={Dubey, Abhimanyu and Jauhri, Abhinav and Pandey, Abhinav and Kadian, Abhishek and Al-Dahle, Ahmad and Letman, Aiesha and Mathur, Akhil and Schelten, Alan and Yang, Amy and Fan, Angela and others},
  journal={arXiv e-prints},
  pages={arXiv--2407},
  year={2024}
}

@article{bi2024deepseek,
  title={Deepseek llm: Scaling open-source language models with longtermism},
  author={Bi, Xiao and Chen, Deli and Chen, Guanting and Chen, Shanhuang and Dai, Damai and Deng, Chengqi and Ding, Honghui and Dong, Kai and Du, Qiushi and Fu, Zhe and others},
  journal={arXiv preprint arXiv:2401.02954},
  year={2024}
}

@article{chiang2023vicuna,
  title={Vicuna: An open-source chatbot impressing gpt-4 with 90\%* chatgpt quality},
  author={Chiang, Wei-Lin and Li, Zhuohan and Lin, Ziqing and Sheng, Ying and Wu, Zhanghao and Zhang, Hao and Zheng, Lianmin and Zhuang, Siyuan and Zhuang, Yonghao and Gonzalez, Joseph E and others},
  journal={See https://vicuna. lmsys. org (accessed 14 April 2023)},
  volume={2},
  number={3},
  pages={6},
  year={2023}
}

@article{touvron2023llama,
  title={Llama 2: Open foundation and fine-tuned chat models},
  author={Touvron, Hugo and Martin, Louis and Stone, Kevin and Albert, Peter and Almahairi, Amjad and Babaei, Yasmine and Bashlykov, Nikolay and Batra, Soumya and Bhargava, Prajjwal and Bhosale, Shruti and others},
  journal={arXiv preprint arXiv:2307.09288},
  year={2023}
}

@article{wang2023openchat,
  title={Openchat: Advancing open-source language models with mixed-quality data},
  author={Wang, Guan and Cheng, Sijie and Zhan, Xianyuan and Li, Xiangang and Song, Sen and Liu, Yang},
  journal={arXiv preprint arXiv:2309.11235},
  year={2023}
}

@inproceedings{shen2024anything,
  title={" do anything now": Characterizing and evaluating in-the-wild jailbreak prompts on large language models},
  author={Shen, Xinyue and Chen, Zeyuan and Backes, Michael and Shen, Yun and Zhang, Yang},
  booktitle={Proceedings of the 2024 on ACM SIGSAC Conference on Computer and Communications Security},
  pages={1671--1685},
  year={2024}
}

@article{chu2024comprehensive,
  title={Comprehensive assessment of jailbreak attacks against llms},
  author={Chu, Junjie and Liu, Yugeng and Yang, Ziqing and Shen, Xinyue and Backes, Michael and Zhang, Yang},
  journal={arXiv e-prints},
  pages={arXiv--2402},
  year={2024}
}

@article{xu2024comprehensive,
  title={A comprehensive study of jailbreak attack versus defense for large language models},
  author={Xu, Zihao and Liu, Yi and Deng, Gelei and Li, Yuekang and Picek, Stjepan},
  journal={arXiv preprint arXiv:2402.13457},
  year={2024}
}


\end{document}